\documentstyle[prabib,pra,aps,epsfig]{revtex}

\begin{document}
\newcommand{\be}{\begin{equation}}
\newcommand{\ee}{\end{equation}}
\newcommand{\bez}{\begin{equation*}}
\newcommand{\eez}{\end{equation*}}
\newcommand{\beqa}{\begin{eqnarray}}
\newcommand{\eeqa}{\end{eqnarray}}
\newcommand{\beqaz}{\begin{eqnarray*}}
\newcommand{\eeqaz}{\end{eqnarray*}}

\draft
\title{An intense, slow and cold beam of metastable Ne(3s)$\, ^3$P$_2$
atoms.}
 \author{J.G.C.\ Tempelaars, R.J.W.\ Stas, P.G.M. Sebel,
H.C.W.\ Beijerinck and E.J.D.\ Vredenbregt}
 \address{Physics Department, Eindhoven University of Technology,
P.O.\ Box 513, 5600 MB Eindhoven, The Netherlands\\E-mail address:
E.J.D.Vredenbregt@TUE.nl}
\date{\today}
\maketitle
 \begin{abstract} We employ laser cooling to intensify and cool an atomic
beam of metastable Ne(3s) atoms. Using several collimators, a slower and
a compressor we achieve a $^{20}$Ne$^*$ flux of $6\times 10^{10}$ atoms/s
in an 0.7~mm diameter beam traveling at 100 m/s, and having longitudinal
and transverse temperatures of $25$mK and $300\mu$K, respectively. This
constitutes the highest flux in a concentrated beam achieved to date with
metastable rare gas atoms. We characterize the action of the various
cooling stages in terms of their influence on the flux, diameter and
divergence of the atomic beam. The brightness and brilliance achieved are
2.1$\times 10^{21}$~s$^{-1}$m$^{-2}$sr$^{-1}$ and 5.0$\times
10^{22}$~s$^{-1}$m$^{-2}$sr$^{-1}$, respectively, comparable to the
highest values reported for alkali-metal beams. Bright beams of the
$^{21}$Ne and $^{22}$Ne isotopes have also been created.
\end{abstract}

\pacs{39.25.+k, 32.80.Lg, 32.80.Pj}

%\twocolumn
\section{Brightening rare-gas atomic beams}

The efficiency with which metastable rare gas atoms can be produced in
gas discharge sources is notoriously low, usually only on the order of
10$^{-5}$ with the highest reported value being 10$^{-3}$ for
helium~\cite{BruHab}. Traditionally, therefore, metastable rare gas
atomic beams have shown much smaller flux and density than alkali-metal
beams. In recent years, several groups
\cite{Hoogerland1,Koolen,Shimizu2,Aspect,Scholz,Nellessen,Schiffer,Engels,Hoogerland2,Hoogerland3,Rooijakkers,Herschbach,Tol,Leduc,Labeyrie,Nowak}
have tried to bridge this gap by employing laser cooling
techniques~\cite{HalPeter} to intensify rare gas beams. At the same time,
atomic beams could be made slow and monochromatic.

The principle of beam intensification using laser manipulation of atomic
trajectories was illustrated by Sheehy {\it et al} in a 1990
paper~\cite{Hal}. The essential element in any such scheme is a laser
collimator, whose purpose is to increase the solid angle under which
atoms can leave the source and still contribute to the beam flux further
downstream. This is achieved by cooling the velocity component transverse
to the atomic beam axis. The collimator completely determines the maximum
gain in beam flux that can be obtained with laser cooling. Without
collimation, the flux of atoms through a constant area decreases
geometrically with its distance from the source; with collimation, this
flux is essentially constant. Collimators for rare gas atomic beams have
been employed by several groups
\cite{Hoogerland1,Koolen,Shimizu2,Aspect,Hoogerland3,Rooijakkers,Herschbach,Tol,Leduc,Nowak}.

Since transverse cooling takes a certain time and thus a certain
transverse distance, laser collimation inevitably leads to large-diameter
atomic beams whose density may be rather low despite the increased beam
flux. The solution to this, according to Ref.~\cite{Hal}, is to focus the
beam and re-collimate it at the focal point. This idea was implemented in
its purest form by Hoogerland {\it et al}~\cite{Hoogerland1} who used a
separate magneto-optical lens (MOL) followed by a second collimator.
Scholz {\it et al}~\cite{Scholz}, Nellissen {\it et al}~\cite{Nellessen}
and Schiffer {\it et al}~\cite{Schiffer} developed a magneto-optical
compressor (MOC) in which both focusing and collimation occur within the
same device. The mayor difference between a MOL and a MOC is the time
spent within the device interacting with the laser fields: a compressor
is basically a lens with a focal length shorter than the length of the
interaction region. In order for the length of the MOC to stay within
reasonable limits, laser slowing of the atomic
beam~\cite{HalPeter,Phillips} between collimator and compressor is
necessary. A MOC device has since been applied to rare gas beams by
Buckman and coworkers~\cite{Hoogerland2,Hoogerland3} and by
Koolen~\cite{Koolen}. Koolen used a lens in combination with a compressor
to increase the spatial capture range of the device. Labeyrie {\it et
al}~\cite{Labeyrie} have also reported on the use of a MOL.

In this paper we describe our solution to brightening rare gas atomic
beams, using all the devices mentioned above extended with additional
cooling stages. This setup stands out because it produces the highest
flux (6$\times 10^{10}$ Ne(3s)$ ^3$P$_2$ atoms/s) in a concentrated beam
of any rare-gas atomic beam setup reported to date, comparable even to the
brightest available alkali-metal beam~\cite{Lison}.

\section{Experimental Setup}

\subsection{Overview}

\begin{figure}
\centering \epsfig{figure=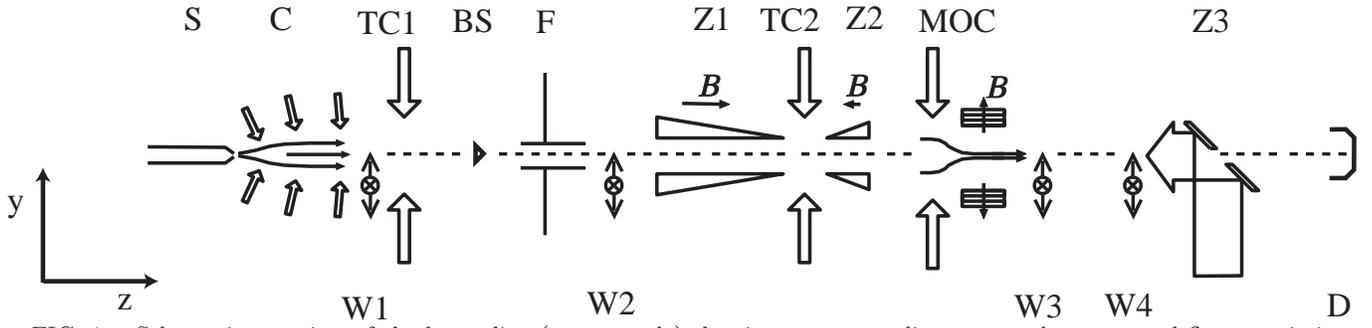,width=18cm}
 \caption{\label{figsetup}
Schematic overview of the beam line (not to scale) showing source,
cooling stages, detectors and flow restriction. S: source, C: collimator,
TC1,TC2, first and second transverse coolers; BS: beam stop (diameter
3~mm), F: flow restriction (150~mm$\times \phi$30~mm, Z1,Z2: tapered coils
of Zeeman slower, Z3: 45$^o$ mirror with 3~mm diameter hole ("slower
mirror"), MOC: magneto-optical compressor, D: beam flux detector,
W1,W2,W3,W4: wire scanners.}
\end{figure}

Figure~\ref{figsetup} shows a schematic overview of the neon beam line,
with five separate laser cooling stages used to create an intense beam of
cold atoms. In all cooling stages we use the
Ne(3s)$\,^3$P$_2\leftrightarrow$(3p)$\,^3$D$_3$ optical transition at a
wavelength $\lambda=640.225$~nm. Important parameters for this transition
are the linewidth $\Gamma=~8.5(2\pi)$MHz, saturation intensity $I_s$=
42~W/m$^2$ for circularly polarized and $I_s$= 72~W/m$^2$ for linearly
polarized light, recoil velocity $v_R = 31$~mm/s and Doppler temperature
$T_D$ = 203$\mu$K~\cite{HalPeter}.

First, a collimator captures metastable Ne(3s) atoms from a discharge
source and collimates them into a parallel beam. Then, an additional
transverse cooling stage reduces the divergence of the beam to a few
times the Doppler limit. In a Zeeman slower, the atoms are axially slowed
to a velocity of $100$~ms$^{-1}$. Again, an additional transverse cooling
stage, positioned in between the two solenoids of the slower, reduces the
divergence of the slowed atomic beam. Behind the slower, a magneto-optical
compressor captures the slowed atoms and funnels them into a narrow beam,
which then passes through a hole in a mirror and hits a detector that
records the beam flux achieved. The total length of the beam line, taken
from the nozzle of the discharge to the beam flux detector, is
approximately $3.3$~m.

Table~I gives detailed information about the position and length
of the different laser cooling stages and the position of
detectors. We take the $z$ direction along the atomic beam axis.

\subsection{Diagnostics}

Four sets of wire scanners are positioned along the beam line for
diagnostics. A wire scanner consists of a stainless steel wire
that can be moved transversely through the atomic beam by a
stepper motor. By scanning the wire through the atomic beam and
measuring the current due to electrons emitted from the wire, a
one-dimensional beam-current profile $I(x)$ is generated. The beam
profile represents a line-integral over the two-dimensional
density distribution $\Phi_m(x,y)$ of the atomic beam along the
length $l$ of the wire,
 \be I(x) \approx \eta_{\rm{A}} e d_w\int_{-l/2}^{l/2}\Phi_m(x,y)dy,\ee
with $\eta_{\rm{A}}$ the Auger detection efficiency of the wire, $e$ the
elementary charge, and $d_w$ the diameter of the wire. Throughout this
paper we use $\eta_{\rm{A}}=1$ since the exact value of the quantum
efficiency of the Auger process for Ne($^3$P$_2$) atoms on metallic
surfaces is not known (Hotop~\cite{Hotop} reports
$\eta_{\rm{A}}=0.3-0.91$ for stainless steel); however, $\eta_{\rm{A}}$
is certainly less than 1 so that in this paper we effectively always
report a lower limit for the atom flux.

The total flux $\dot{N}$ of metastable atoms in the beam at the position
of a wire scanner is found by integrating $I(x)$ over the scan direction
$x$. By comparing beam profiles taken with two successive scanners,
information about the divergence of the atomic beam can be obtained. We
calculate the divergence $\Theta$ of the atomic beam by dividing the
difference in FWHM beam diameters $d_i$ measured with the two wire
scanners by their separation, $\Theta=(d_2-d_1)/(z_2-z_1)$.

Finally, on the way to a trapping chamber where the bright beam can be
used, e.g., for collision experiments or to load a magneto-optical
trap~\cite{Kuppens}, the atomic beam hits a conducting surface that can
be moved in and out of the beam's path, which we use to measure the total
useful atom flux.

\subsection{Laser setup}

All laser cooling stages are operated from a single continuous-wave
single-frequency ring dye laser (Coherent 899-21) with a maximum output
power of $\approx 1$~W when pumped with $10$~W of light from an Argon ion
laser (Coherent Innova 315). We typically set the dye laser output to
700~mW; this is the output power used for the experiments reported here
unless explicitly noted otherwise. The laser is locked to a frequency that
is shifted from the Ne (3s) $^3$P$_2 \leftrightarrow$ (3p) $^3$D$_3$
transition by using Zeeman-tuned saturated absorption spectroscopy. This
"global detuning" $\Delta_L$ of the dye laser is normally set at
$\Delta_L=-1.8\Gamma$ for reasons that will become apparent below. Beams
for the various cooling stages are split off by combinations of half-wave
plates and polarizing beam-splitter cubes. Two acousto-optic modulators
(AOMs) are used to shift the laser frequency for the collimator (80 MHz
AOM) and the Zeeman slower (400 MHz). Spherical and cylindrical
telescopes are used to expand the laser beams to the required sizes;
generally we expand a laser beam such that its FWHM diameters in the $x$-
and $y$-directions approximately match the spatial extent of a cooling
stage's interaction region; in this case the edges of the interaction
region are irradiated at half the central intensity. Then, 50\% of the
power falls within the interaction area and the average intensity
$\langle I \rangle$ is 72\% of the peak value.

Table~I gives details on the laser beam characteristics of each cooling
stage.

\subsection{Metastable atom source}

The beam line starts with a liquid nitrogen cooled source that produces a
beam of metastable $^{20}$Ne(3s)$\,^3$P$_2$ atoms in a DC discharge that
runs through the nozzle of a supersonic expansion. The average axial
velocity of the atoms is $480$~m/s with a FWHM width of 100~m/s
(corresponding to a source temperature of 200~K and a longitudinal beam
temperature of 4~K); the center-line intensity for $^{20}$Ne(3s)$^3$P$_2$
atoms is $2\times 10^{14}$~s$^{-1}$sr$^{-1}$. Other high-energy products
emerging from the source are atoms in the metastable $^3$P$_0$ state,
metastable atoms of the isotopes $^{21}$Ne and $^{22}$Ne, and UV photons.

\subsection{Collimator}

About 40~mm after the source, the atoms enter the collimator in which they
are captured and collimated by a two-dimensional optical
molasses~\cite{HalPeter}. For the molasses beams we use curved wave
fronts to achieve a large capture angle while keeping the cooling time as
short as possible. To minimize the required laser power, the curved wave
fronts are produced by using the zig-zag method as we reported
earlier~\cite{Hoogerland1}. Linearly polarized laser light is injected at
an angle $\beta(\Delta z=0)$ with respect to the plane perpendicular to
the atomic beam axis between two 150~mm long, nearly parallel mirrors
placed 60~mm apart. With each reflection, $\beta(\Delta z)$ is reduced by
an amount $\alpha=0.85$~mrad, half the angle between the mirrors. The
Doppler shift $\Delta_D$ experienced by the atoms in the collimator is
given by
 \be \Delta_D=-\textbf{k}\cdot\textbf{v}
 \approx -kv_{||}\beta +kv_{\perp}, \ee
with $v_{||}$ and $v_{\perp}$ the longitudinal and transverse
velocity of the atoms. With the condition $\Delta_D+\Delta_L=0$
the resonant transverse velocity of the atoms is
 \be v_{\perp}(\Delta z)=-\Delta_L/k+v_{||}\beta(\Delta z).\label{effdetuning} \ee
At the entrance of the collimator this velocity is $v_{\perp}(\Delta
z=0)=21$~ms$^{-1}$ for the experimental parameters we use, i.e.,
$\beta(\Delta z=0)=130$~mrad and $\Delta_L=+65$~$(2\pi)$MHz, resulting in
a FWHM capture angle $\Theta_c=2 v_{\perp}/\bar{v}_{||}=86$~mrad. At the
end of the collimator at $\Delta z=150$~mm, i.e., after 12 reflections of
the laser beam on each mirror, the resonance velocity is reduced to
$v_{\perp}(\Delta z=150$~mm$)=11$~m/s. Clearly, this effective detuning
at the end of the collimator is too large to provide the lowest possible
divergence of the atomic beam; in addition it depends on the axial
velocity of the atoms. Therefore, an additional, short, transverse
cooling stage is positioned immediately behind the collimator to further
reduce the divergence of the atomic beam. Here we use $\Delta =
-1.8\Gamma$ and $\beta$=0; in-vacuum mirrors create the two-dimensional
molasses from a single, linearly polarized laser beam.

\begin{figure}
\centering \epsfig{figure=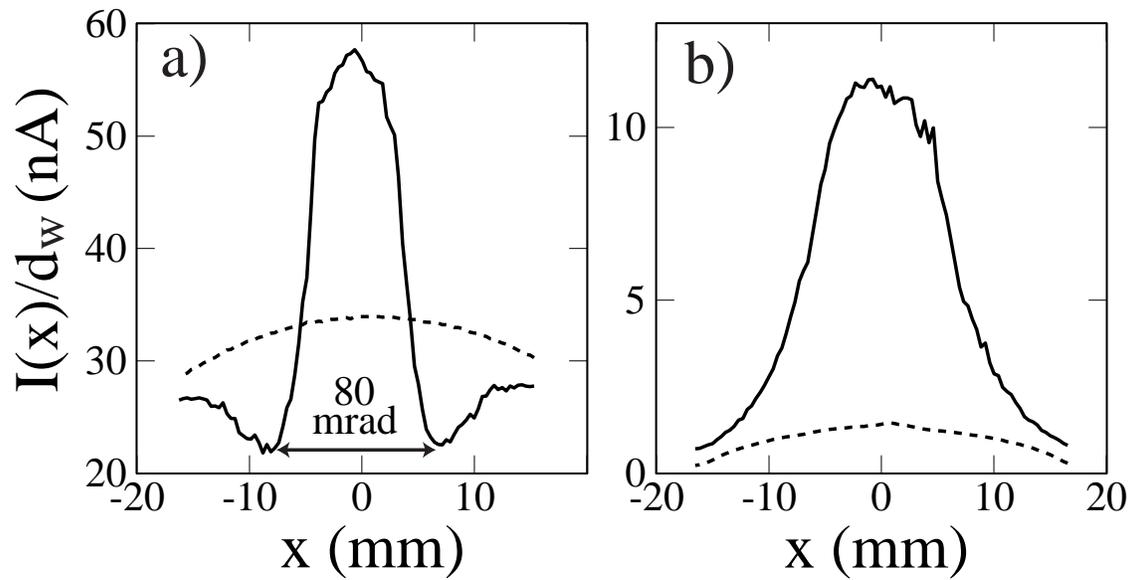,width=15cm}
 \caption{\label{figprofcoll}
Experimental spatial beam profiles behind collimating stage taken (a)
just behind the collimator with wire scanner W1 and (b) 580~mm further
downstream with wire scanner W2. Solid lines: cooling stages on; dashed
lines: cooling stages off. The apparent loss in flux for the dashed lines
between Fig.~(a) and (b) is due to the influence of the flow resistance
(see Fig.~\protect{\ref{figsetup}}).}
\end{figure}

Figure~\ref{figprofcoll} shows experimental beam profiles taken in the
$x$ direction with wire scanner W1 (which is placed just behind the
collimator), and with scanner W2 placed $580$~mm further downstream. The
effect of the collimator is clearly visible: the atoms are captured and
cooled in transverse direction, resulting in an atomic beam with a FWHM
diameter of $d_{\rm{col}}=9$~mm containing $\dot{{\rm N}}_{{\rm
col}}=1.1\times 10^{12}$ atoms/s in the $^3$P$_2$ state. From the figure
the capture angle of the collimator is estimated at $\Theta_c=80$~mrad,
approximately the same as found with Eq.~(\ref{effdetuning}). The
measurements with the second scanner then show a FWHM divergence of the
beam of $10$~mrad. This value could in principle be reduced further by
changing the detuning of the transverse cooling stage; the detuning
chosen is a compromise that leads to maximum overall output of the beam
machine but which does not necessarily optimize each stage individually
(we discuss this further in section~III).

Switching the collimation and transverse cooling laser off, the beam flux
decreases drastically, as shown by the dashed line in
Fig.~\ref{figprofcoll}b. This beam profile contains, besides atoms in the
$^3$P$_2$ state, also unwanted products of the source, i.e., atoms in the
metastable $^3$P$_0$ state, metastable atoms of the other isotopes, and
UV photons. Therefore a beam stop, consisting of a disc with a diameter
of $3$~mm, can be positioned in the center of the atomic beam just behind
the transverse cooler (see Fig.\ 1). This beam stop keeps unwanted
products (including ground state atoms) from reaching the end of the
setup, while most of the laser cooled atoms pass around it, due to the
different nature of the trajectories of cooled atoms versus uncooled
atoms and UV photons.

\subsection{Zeeman slower} \label{zeeh3}

The collimated atoms then enter a midfield-zero Zeeman
slower~\cite{HalPeter,Phillips} in which they are decelerated by a
counter-propagating laser beam. The laser beam is coupled in by a mirror
positioned in the vacuum behind the magneto-optical compressor (Z3 in
Fig.~\ref{figsetup}). The atomic beam can pass through a 3~mm diameter
hole in the center of this "slower mirror" since the compressor greatly
reduces the beam's radius. The length of the slower is such that 10\% of
the maximum possible scattering force is sufficient to slow the atoms
down. The tapered magnetic field coils (0.85~m and 0.15~m long) are
usually set to produce maximum fields of 275~Gauss and -175~Gauss. In
combination with the 400~MHz detuning of the circularly polarized slowing
laser, this means that atoms with initial velocities up to 500~m/s should
be slowed to a final velocity close to 100~m/s.

\begin{figure}
\centering \epsfig{figure=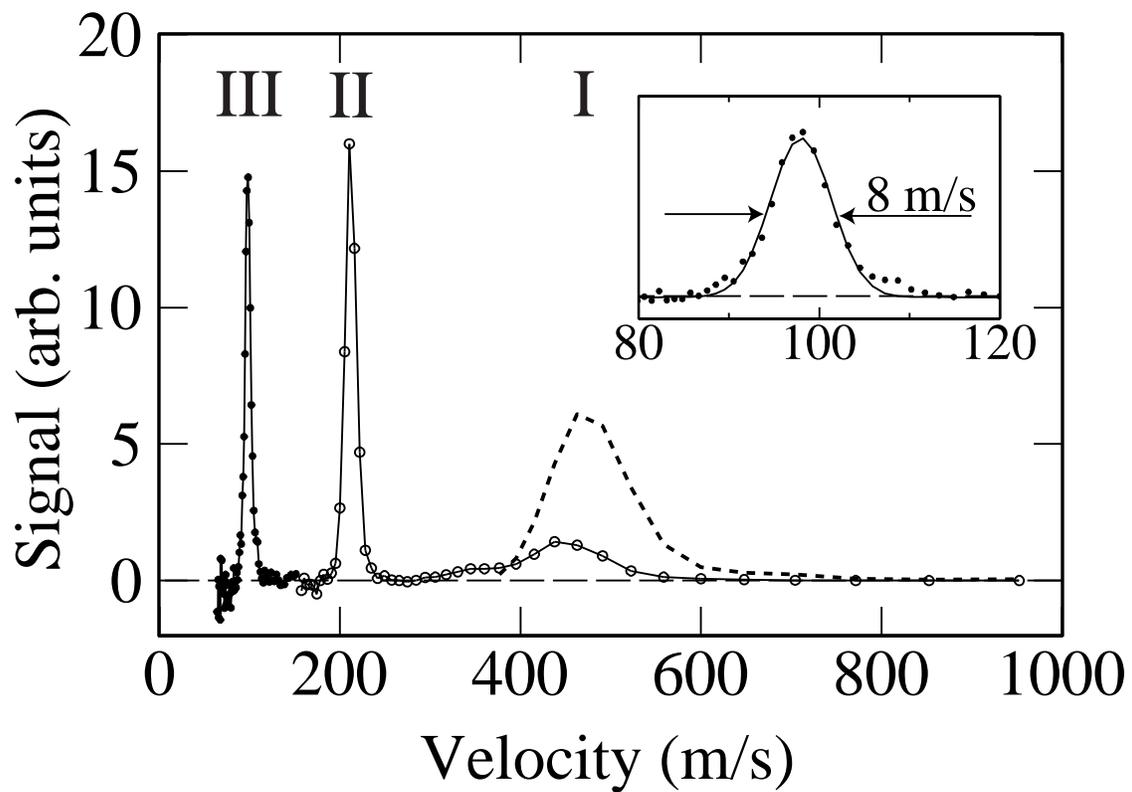,width=15cm}
 \caption{\label{figveldist}
Measured time-of-flight velocity distributions for unslowed (I),
partially slowed (II) and fully slowed atoms (III). The inset shows a
Gaussian fit to the peak of fully slowed atoms, resulting in an 8.0~m/s
FWHM velocity width.}
\end{figure}

We measured the longitudinal velocity of the atoms leaving the Zeeman
slower by a standard time-of-flight technique. Figure~\ref{figveldist}
shows the measured velocity distribution of unslowed, partially slowed
(second Zeeman solenoid operated with reduced current), and fully slowed
atoms. As expected, fully slowed atoms have a final velocity of 98~m/s.
The measured FWHM width is 8.0~m/s in this case, corresponding to a
longitudinal beam temperature of $T_{||}=25$~mK.

During the slowing process the divergence of the atomic beam increases
due to the reduction in axial velocity as well as the randomness in the
direction of the spontaneously emitted photons. To counteract this
effect, an additional transverse cooling stage is inserted between the
two solenoids of the Zeeman slower. Transverse cooling is possible here
because the magnetic field vanishes between the two solenoids. Here we
use a simple "V"-shaped arrangement of two in-vacuum mirrors to create the
two-dimensional molasses from a single, linearly polarized laser beam.
Reducing the divergence of the atomic beam increases the flux of atoms
within the capture range of the magneto optical compressor. We obtain a
factor two increase in beam flux behind the MOC in this way.

\subsection{Magneto-optical compressor}

\begin{figure}
\centering \epsfig{figure=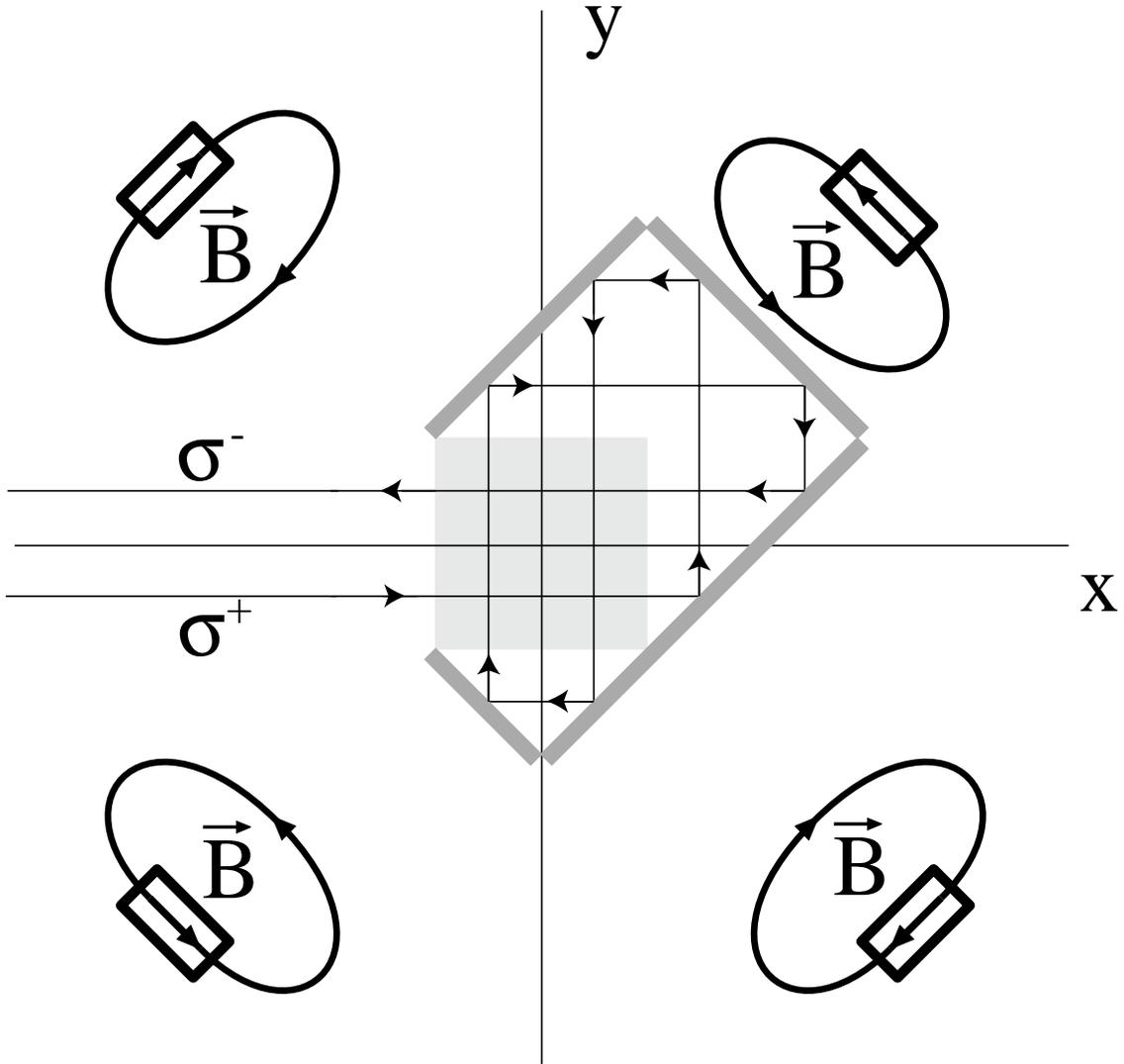,width=15cm}
 \caption{\label{figmocview}
Schematic cross section of the magneto-optical compressor. In-vacuum
mirrors create two pairs of $\sigma^+/\sigma^-$ molasses beams. Four
NdFeB magnets, positioned near the end of the mirrors, are arranged to
create a quadrupole magnetic field whose strength increases along $z$.}
\end{figure}

Behind the Zeeman slower, the atomic beam passes through a magneto-optic
compressor to reduce its diameter and divergence. Figure~\ref{figmocview}
shows a schematic view of the device, which is basically a
two-dimensional magneto-optic trap where the magnetic field gradient
increases along the symmetry axis. Scholz {\it et al}~\cite{Scholz} have
given a description of the motion of atoms inside a MOC.

Two pairs of $\sigma^+-\sigma^-$ laser beams are produced by bouncing a
single $\sigma^+$ laser beam on a set of in-vacuum mirrors. Atoms
interact over a distance of $2w_{z}=90$~mm with the laser beams, whose
average on-resonance saturation parameter is $\langle s_0 \rangle=\langle
I\rangle/I_s=1.9$ while the detuning $\Delta_L=-1.8\Gamma$.

The magnetic quadrupole field of the MOC is generated by four permanent
magnets mounted outside of the vacuum system near the end of the mirror
arrangement. The NdFeB magnets have dimensions
60$\times$60$\times$30~mm$^3$ and a magnetization of 1.15~T. At the
entrance of the MOC a $\mu$-metal shield helps to reduce the quadrupole
field gradient to $0.1$~T/m, which increases to $0.6$~T/m at the end of
the MOC. Given the initial field gradient and laser detuning, an initial
resonance circle with radius $r_c= -\Delta_L/(\mu_B \nabla{B})$=10~mm
exists within which all atoms will be captured and funneled to the beam
axis. Here, $\mu_B$ is the Bohr magneton.

\begin{figure}
\centering \epsfig{figure=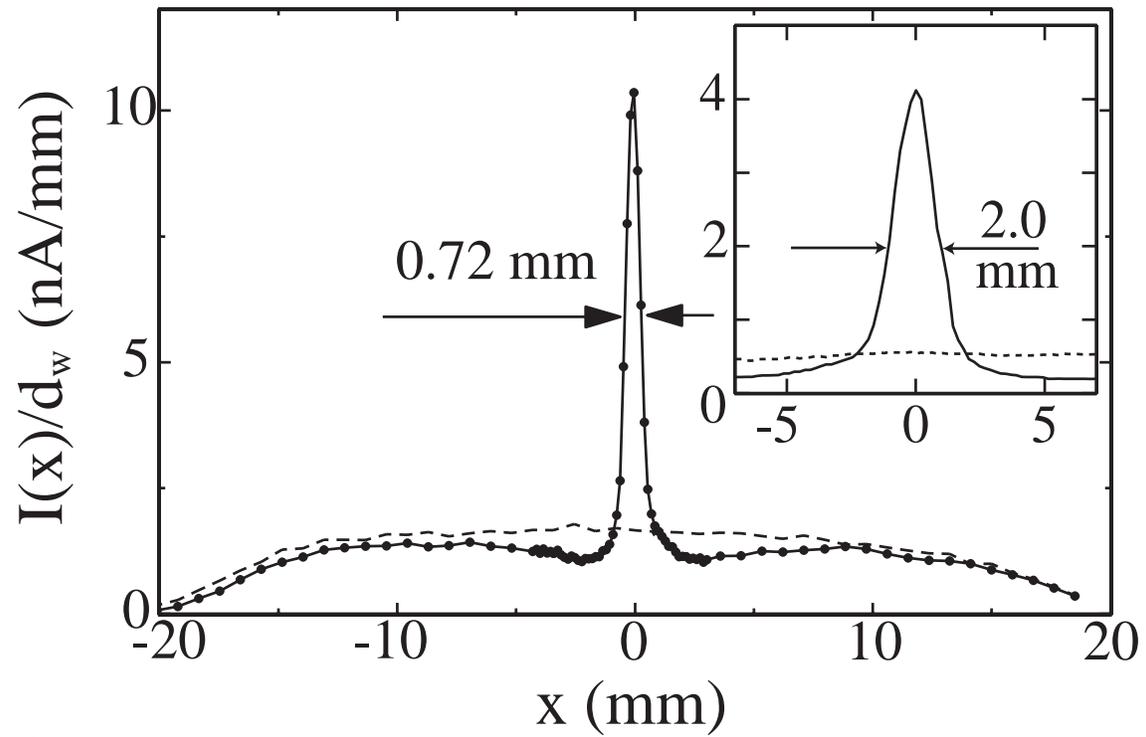,width=15cm}
 \caption{\label{figprofcomp}
Spatial beam profile measured immediately behind the compressor with wire
scanner W3, showing a beam diameter $d=0.72$~mm. The inset shows the
profile 150~mm further downstream, measured with wire scanner W4. Solid
lines, solid circles: compressor on; dashed line: compressor off.}
\end{figure}

Figure~\ref{figprofcomp} shows beam profiles measured with the vertical
wire of the third and fourth wire scanners. The effect of the MOC is very
visible: atoms are molded into a narrow beam with a diameter of
$d_{\rm{MOC}}=0.72$~mm and a divergence of 8~mrad, containing
$\dot{N}_{\rm{MOC}}=5\times 10^{10}$~atoms/s. This results in a particle
density of $n=4\dot{N}/(\pi d^2 \bar{v}_{||})=1.2\times 10^{9}$~cm$^{-3}$.
From the figure it is also clear that many of the slowed atoms arrive
outside the capture range of the MOC and cannot contribute to the flux of
the compressed beam.

The measured FWHM divergence $\Delta v_\perp = 0.84$ m/s approximately
corresponds to the FWHM transverse Doppler velocity $\Delta v_D =0.68$
m/s, resulting in a transverse beam temperature $T_{\perp}=300$~$\mu$K.
One might expect the transverse temperature to be lower than this value,
since sub-Doppler cooling is possible when atoms are sufficiently near
the beam axis at the end of the MOC~\cite{Schiffer}. We have found,
however, that the final divergence is limited to approximately the
Doppler value by a Stern-Gerlach effect. After the atoms exit the
compressor's interaction region, they still have to traverse a large
quadrupole field which exerts a force~\cite{Koolen}
 \be \vec{F}_{{\rm{SG}}}=(\vec{\mu}\cdot\nabla)\vec{B}=-\mu_B g_g m_g
 \vec{\nabla} |B|, \ee
directed transverse to the beam axis, leading to motion away from the
axis (this Stern-Gerlach force can be ignored inside the MOC because
there the optical forces dominate). Here, $m_g$ is the magnetic sublevel
occupied by the atom when it leaves the compressor's interaction region
and $g_g$=1.5 the Land\'{e} factor. We calculate that in our case
Stern-Gerlach forces contribute around 4~mrad to the divergence of the
atomic beam~\cite{Stas}, effectively ruling out sub-Doppler transverse
collimation. (Of course, further collimation is possible once outside the
range of the quadrupole field.)

\section{Overall performance}

\subsection{Detuning and intensity dependence}

\begin{figure}
\centering \epsfig{figure=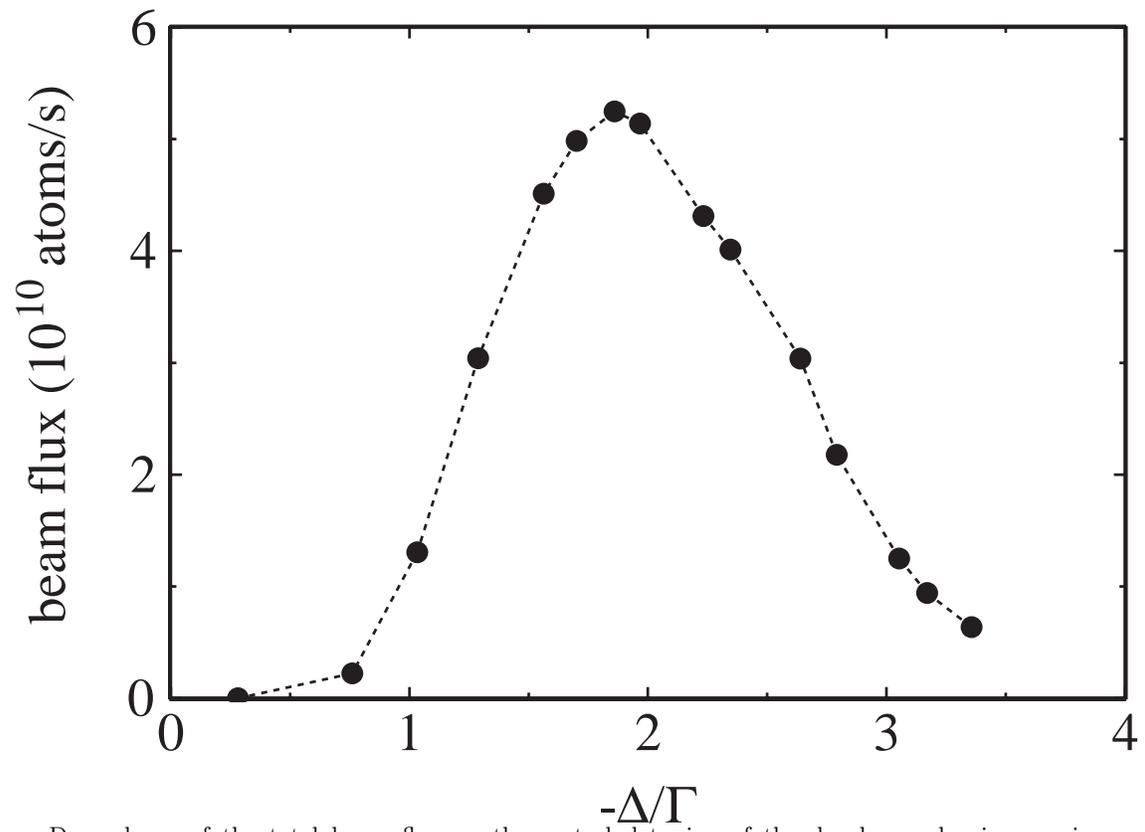,width=15cm}
 \caption{\label{fignvsd}
Dependence of the total beam flux on the central detuning of the
dye laser, showing maximum output at $\Delta_L=-1.8\Gamma$.}
\end{figure}

As mentioned earlier, the detuning used in the various stages is the
result of a compromise that results from deriving all laser beams from a
single source using a very limited amount of AOMs. In Fig.~\ref{fignvsd}
we show the dependence of the beam flux as measured behind the slower
mirror on the overall detuning of the dye laser. A fairly sharp maximum
appears at $\Delta=-1.8\Gamma$, which was consequently chosen as the
working point for the setup.

\begin{figure}
\centering \epsfig{figure=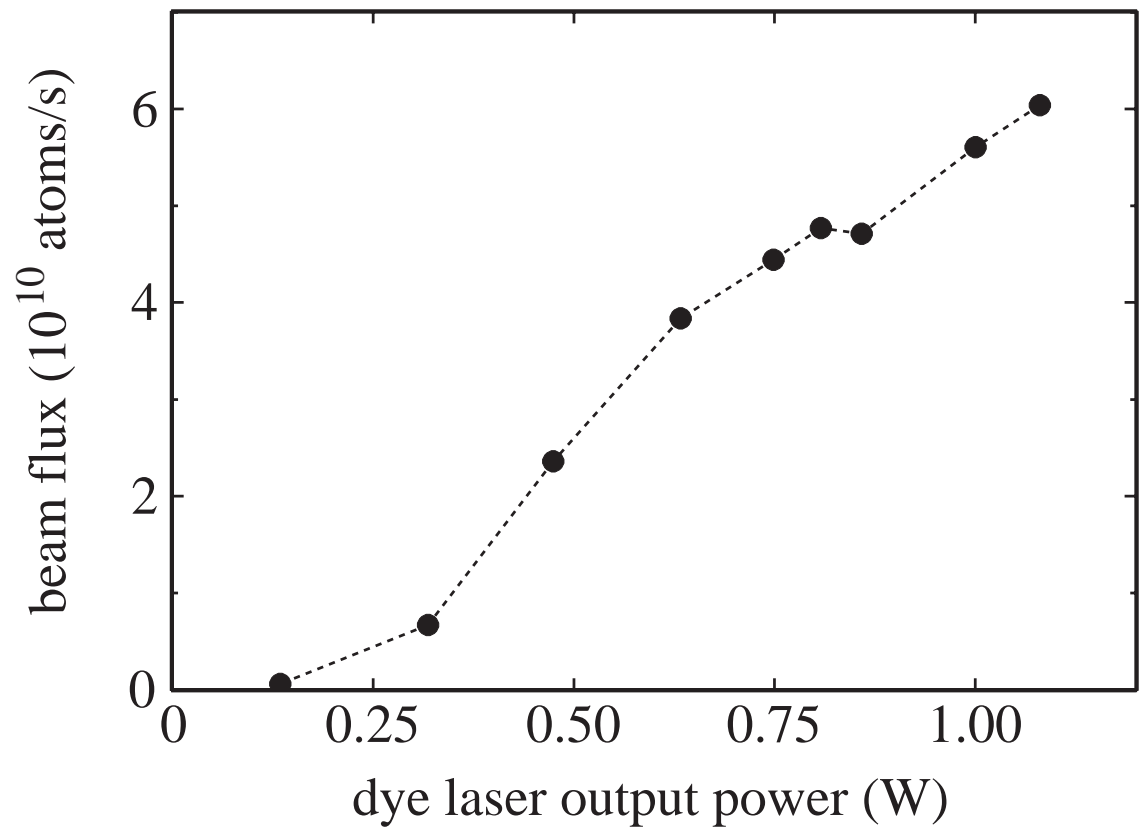,width=15cm}
 \caption{\label{fignvsp}
Dependence of the total beam flux on the output power of the dye laser at
a central detuning $\Delta=-1.8\Gamma$.}
\end{figure}

A similar compromise is necessary in order to fix the distribution of
optical power over the various cooling stages. In general, all of these
show increased performance with the on-resonance saturation parameter
$s_0$ up to at least $s_0=2$, requiring more than the nominal output of
the dye laser to attain everywhere. The distribution of power apparent in
Table I is therefore the result of much empirical work.
Figure~\ref{fignvsp} shows the dependence of the output of the beam
machine on the total laser power available from the dye laser. The figure
shows that the output of the setup increases with laser power in the
entire range accessible to us. Consequently, we achieve the maximum beam
flux of $6\times 10^{10}$ metastable $^{20}$Ne atom/s at the maximum
available laser output power of 1W.

\subsection{Brightness and brilliance} In the previous sections we
showed that, by using several sets of wire scanners, it is possible to
measure the characteristics of the atomic beam such as the flux $\dot{N}$
of atoms, the beam diameter and divergence. These quantities can be
summarized into values for the {brightness} ${\cal{R}}$, {brilliance}
${\cal{B}}$ and reduced phase-space density $\tilde{\Lambda}$ of the
atomic beam. Following Lison et al.~\cite{Lison},
 \be {\cal{R}}=\frac{4\dot{N}}{\pi d^2 \Omega},\ee
 \be {\cal{B}}={\cal{R}}\frac{2\bar{v}_{||}}{\Delta v_{||}},\ee
 \be {\tilde{\Lambda}}={\cal{B}}\frac{\pi h^3}{m^3 v_{||}^4},\ee
where $\Delta v_{||}$ is the longitudinal velocity spread and $\Omega$
the geometric solid angle occupied by the beam, $\Omega=\pi(\Delta
v_{\perp}/2\bar{v}_{||})^2$, with $\Delta v_{\perp}$ and ${v}_{||}$ the
FWHM transverse velocity spread and average longitudinal velocity of the
atoms, respectively~\cite{twosnote}.

Table~II lists measured beam characteristics such as beam flux, beam
diameter, longitudinal and transverse velocities, as well as {\it local}
values of the brightness, brilliance and reduced phase space density
calculated from these. Looking at the measured beam flux, going from the
collimator to the end of the setup, about a factor of 20 in beam flux is
lost. This is mostly caused by the fact that only a small fraction of the
slowed atoms can be captured by the magneto-optical compressor, due to its
limited spatial capture range in combination with the rather large
diameter and divergence of the atomic beam leaving the slower. This effect
could be remedied somewhat by using an additional MOL between slower and
MOC~\cite{Koolen}. Also, experiments show that the slower only slows a
fraction of $\approx$ 50\% of the atoms that enter it to 100~m/s,
probably due to imperfections in the slowing laser's beam profile as well
as imperfect optical pumping into the $m_g = 2$ magnetic sub-level at the
start of the slowing process.

The brightness and brilliance of the atomic beam increase drastically
going from source to compressor. This is clear from the last three
columns of Table~II, which give the values of the brightness, brilliance
and reduced phase space density at the end of the setup for different
operation modes of the atomic beam machine, going from limited operation
(source only) to full operation (all laser cooling stages on). Switching
on the slower greatly decreases brightness and brilliance due to the
increasing beam diameter and divergence; the magneto-optical compressor
compensates for this loss. Comparing the operation with only the source
on to the full operation mode, we see that the brightness and brilliance
increase six to seven orders of magnitude and the phase-space density nine
orders of magnitude. Regrettably, the latter is still seven orders of
magnitude from the quantum limit ${\tilde{\Lambda}}=1$.

\subsection{Isotope dependence}

\begin{figure}
\centering \epsfig{figure=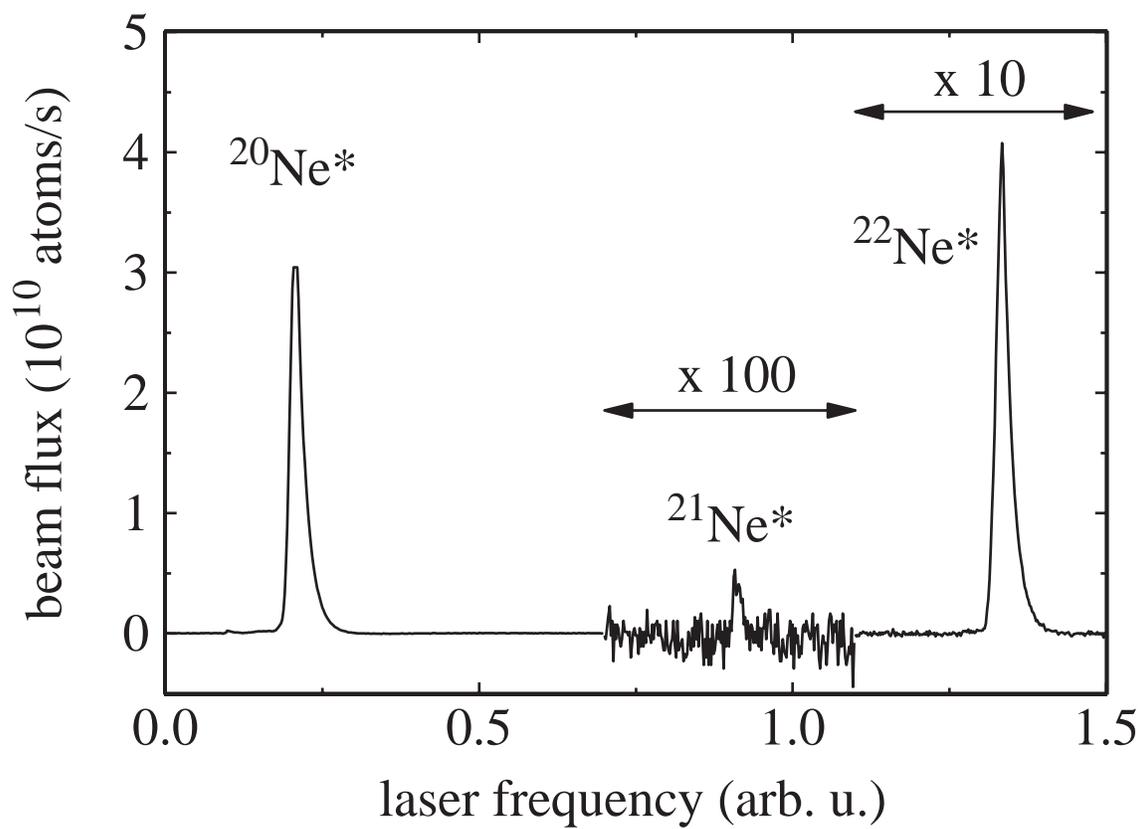,width=15cm}
 \caption{\label{figisotopes}
Total beam flux recorded during $\approx$2GHz wide scan of dye laser
frequency. All three stable isotopes of neon are observed with relative
rates of $^{20}$Ne$^*$ : $^{21}$Ne$^*$ : $^{22}$Ne$^*$ = 1 : 0.0011 :
0.12. (Note dual scale change on vertical axis.)}
\end{figure}

Besides $^{20}$Ne there exist two other stable isotopes of neon, the
bosonic $^{22}$Ne and the fermionic $^{21}$Ne, with natural abundance
ratios of $^{20}$Ne : $^{21}$Ne : $^{22}$Ne = 1 : 0.0030 :
0.102~\cite{rubberbible}. We investigated the achievable beam flux of each
of these by scanning the dye laser over a 2GHz frequency range covering
resonance lines of every isotope~\cite{Laloe} and recording the resulting
beam flux. Figure~\ref{figisotopes} shows the observed signal, with all of
the isotopes indeed appearing. From the areas under the appropriate peaks
we conclude that the ratios of maximum beam flux are $^{20}$Ne$^*$ :
$^{21}$Ne$^*$ : $^{22}$Ne$^*$ = 1 : (0.0011 $\pm$ 0.0002) :
(0.12$\pm$0.01), not far from their natural abundance ratios. The
relatively low output of $^{21}$Ne is easily explained by the fact that,
due to its hyperfine structure, only part of such atoms can be
manipulated at one time (in fact, only the $F=7/2$ state can easily be
laser cooled), while in addition laser cooling for this isotope leads to
losses when there is no repumper laser present~\cite{HalPeter}. In the
light of earlier experiments by Shimizu {\it et al}~\cite{Shimizu1}, who
observed that they could not trap $^{21}$Ne$^*$ unless they supplied
repumper light, the beam flux we achieve without it is somewhat
remarkable. Given the maximum output we obtained with $^{20}$Ne, the
signals recorded imply that under optimum conditions the setup can
deliver a flux of $7\times 10^7$ metastable $^{21}$Ne atoms/s. In our
experience, this is quite sufficient for loading a magneto-optical trap
with around $10^8$ fermionic metastable atoms~\cite{Kuppens}. Of course,
this value could easily be increased by using an enriched gas.

\begin{figure}
\centering \epsfig{figure=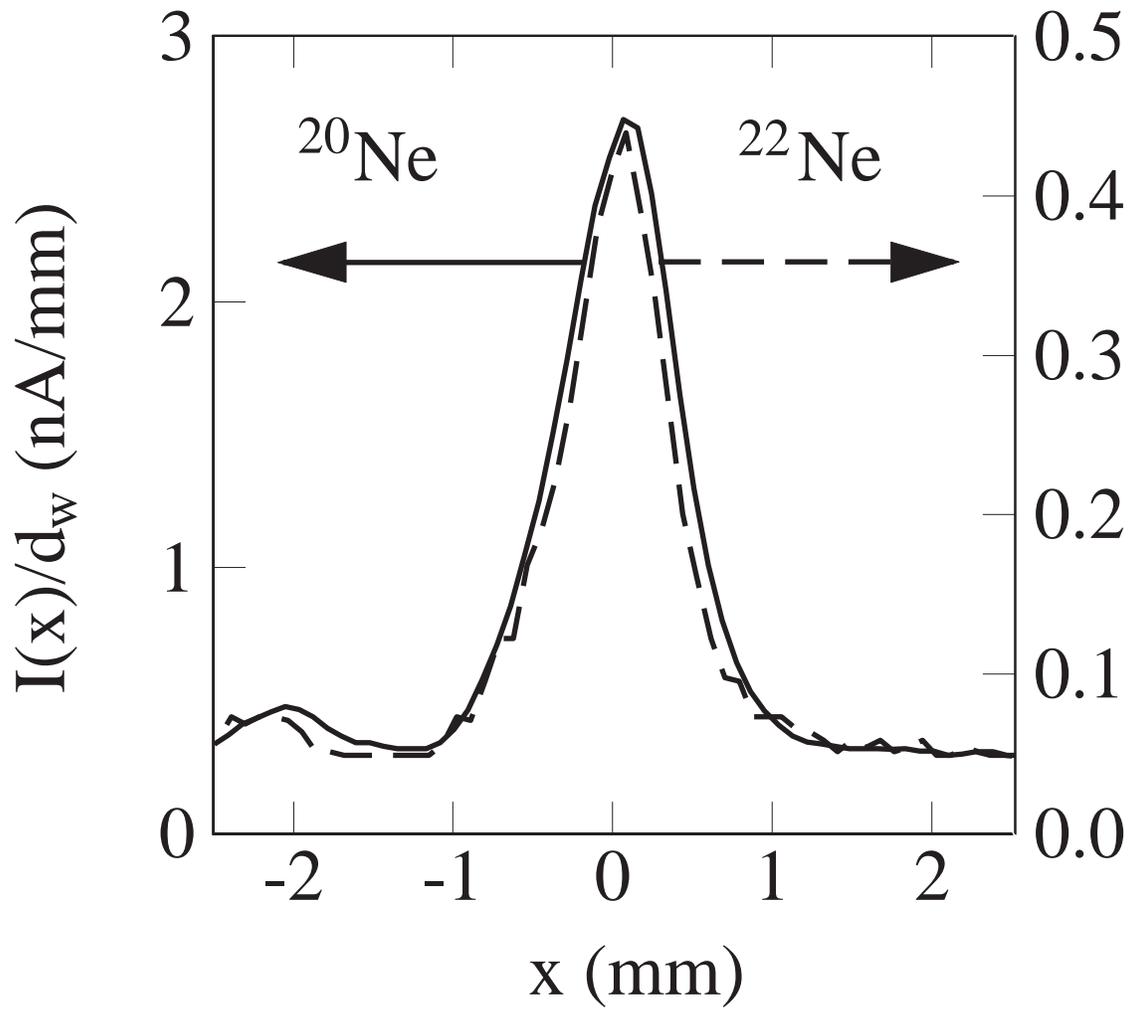,width=15cm}
\caption{\label{figprofiso} Spatial beam profiles for $^{20}$Ne$^*$
(solid line, left vertical axis) and $^{22}$Ne$^*$ (dashed line, right
vertical axis) taken with wire scanner W4 at a central detuning
$\Delta=-1.8\Gamma$.}
\end{figure}

By locking the dye-laser on a saturated absorption signal for $^{22}$Ne,
we studied the spatial profile of the cold and intense beam of this
isotope. As seen in Fig.~\ref{figprofiso}, which shows profiles for both
$^{22}$Ne and $^{20}$Ne measured with the fourth wire scanner, the
$^{22}$Ne atomic beam has virtually the same width and divergence as the
$^{20}$Ne beam, and simply contains about seven times fewer atoms than
the $^{20}$Ne beam.

\section{Discussion and conclusions}

We have described the implementation of an extensive beam brightening
scheme based on manipulation of atomic trajectories by radiation pressure
forces. This technique is especially useful for metastable rare gases
because of the low efficiency with which these species can be produced in
discharge sources. Our setup contains three elements that are essential
for reaching very high brilliance: (1) a collimator that increases the
atom flux, (2) a slower that reduces both the average velocity and the
velocity spread, and (3) a compressor that increases the density by
reducing the beam diameter.

The use of each one of these elements with metastable rare gas beams has
been reported earlier. In most of these cases, however, emphasis has been
on reaching a high flux of slow atoms in order to load a large number of
atoms into a magneto-optical trap (MOT). For such experiments, high
brightness is not essential so that only a collimator and slower suffice.
This combination does not create an atomic beam in the traditional sense
of having a small diameter and small divergence, i.e., having high
brightness. Precise data on useful flux or brightness produced after
slowing have not been published. In this case one could compare MOT
loading rates, but this is somewhat out of the scope of the current paper.

In a separate paper~\cite{Kuppens}, however, we will report on loading a
MOT with our bright beam. Its small divergence allows us to pass the beam
through two 8~mm diameter flow restrictions for differential pumping
before entering the trapping chamber, ensuring very low background
pressure. The low velocity of the beam makes it possible to slow the
atoms down to the capture velocity of the MOT with a $\approx$20~mm long
Zeeman slower positioned immediately before the trap. In combination with
the small diameter of the beam, this results in a very high flux of atoms
delivered directly within the spatial capture range of the MOT, which was
limited to a diameter of 18~mm by the size of the trapping beams that we
used. We have observed MOT loading rates of up to 3.6$\times 10^{10}$
atoms/s, constituting nearly 65\% of the total beam flux. Because of this
high efficiency, we are able to trap up to 9$\times 10^{9}$ metastable
$^{20}$Ne atoms and 3$\times 10^{9}$ metastable $^{22}$Ne atoms in our
MOT under otherwise conventional trapping conditions. These numbers are
substantially higher than reported by
others~\cite{Shimizu2,Herschbach,Robert,Pereira3} for brightened but not
compressed metastable atomic beams.

Applications where high brightness as afforded by the use of a MOC is
clearly profitable, e.g., atom lithography, certain quantum optics
experiments and loading hollow fiber guides, have been reported by Engels
{\it et al}~\cite{Engels}, Koolen~\cite{Koolen} and Dall {\it et
al}~\cite{Hoogerland3,Hoogerland4}, respectively. In the latter two cases,
the experimental setups used are based on virtually the same ideas as
ours but use helium as the rare gas. The difficulties of manipulating
helium, with its comparatively small linewidth, have led to beam fluxes
that are smaller by a factor of 6 to
10~\cite{Koolen,Hoogerland3,Hoogerland4}. Dall {\it et al} report a final
beam diameter of around 5~mm, suggesting a much smaller brightness was
achieved but final data on this is lacking. Koolen's setup produces a
brightness of 1.0$\times 10^{21}$~s$^{-1}$m$^{-2}$sr$^{-1}$, quite
comparable to our result. Engels {\it et al}, who also use neon, report a
beam flux of $6\times 10^8$ atoms/s, which is two orders of magnitude
less than our current values, mostly due to the use of a less efficient
collimator. Since, however, these authors achieved a much smaller beam
diameter ($70\mu$m) than we do as well as somewhat smaller divergence,
their setup delivers a factor 3.3 higher phase space density and a factor
2.3 higher brilliance.

For alkali-metal beams, the highest brightness, 6.3$\times 10^{21}$~Cs
atoms~s$^{-1}$m$^{-2}$sr$^{-1}$ has been reported by Lison {\it et
al}~\cite{Lison}. Their setup uses a combination of a collimator, a slower
and a lens plus re-collimator to produce a flux of $2.6\times 10^{10}$ Cs
atoms/s. Comparing these numbers to our own, we find that they are nearly
the same, notwithstanding the fact that we are forced to start out with a
much lower brightness atom source. This again points out that very large
gains in output are possible with beam brightening techniques which
depend to a large degree on the efficiency of the collimator used.

Finally, there is still room for improvement of our results. For one, the
use of a magneto-optic lens between slower and compressor could increase
the final beam flux by a factor of $\approx 10$ by effectively increasing
the capture range of the compressor~\cite{Koolen}. Also, outside of the
region of strong Stern-Gerlach forces, sub-Doppler transverse cooling of
the bright beam is possible in principle. Given the limits of such
cooling mechanisms~\cite{HalPeter}, a FWHM divergence of
$\approx$2.4~mrad can be achieved leading to a gain in both brightness and
brilliance of another factor of 10.

\section{Acknowledgments}
We are indebted to J.T.M.\ van Beek, W.J.\ Mestrom, V.P.\ Mogendorff,
S.J.M. Kuppens, B.J.\ Claessens, and E.\ van Kempen for help with some of
the experiments, to K.A.H.\ van Leeuwen for many helpful discussions, and
to M.J.\ de Koning, L.A.H.M.\ van Moll and J.A.C.M.\ van de Ven for
technical assistance. We thank A.E.A.\ Koolen for pointing out the
effects of the Stern-Gerlach force to us. This work was supported by the
Netherlands organization for Fundamental Research on Matter (FOM) and by
the Royal Netherlands Academy of Arts and Sciences (KNAW).

\bibliographystyle{prsty}
%\bibliography{refs}

\begin{thebibliography}{10}

\bibitem{BruHab}
B. Brutschy and H. Haberland, J. Phys. E {\bf 10},  90  (1977).

\bibitem{Hoogerland1}
M. Hoogerland {\it et~al.}, Appl. Phys. B {\bf 62},  323  (1996).

\bibitem{Koolen}
A. Koolen, Ph.D. thesis, Eindhoven University of Technology, (2000); see
also R.M.S. Knops {\it et al}, Laser Phys. {\bf 9} (1999) 286.

\bibitem{Shimizu2}
F. Shimizu, K. Shimizu, and H. Takuma, Chem. Phys. {\bf 145},  327 (1990)

\bibitem{Aspect}
A. Aspect {\it et~al.}, Chem. Phys. {\bf 145},  307  (1990)

\bibitem{Scholz}
A. Scholz {\it et~al.}, Opt. Commun. {\bf 111},  155  (1994)

\bibitem{Nellessen}
J. Nellessen, J. Werner, and W. Ertmer, Opt. Commun. {\bf 78},  300
(1990).

\bibitem{Schiffer}
M. Schiffer, M. Christ, G. Wokurka, and W. Ertmer, Opt. Commun. {\bf
134},  423 (1997)

\bibitem{Engels}
P. Engels {\it et~al.}, Appl. Phys. B {\bf 69},  407  (1999)

\bibitem{Hoogerland2}
D. Milic, M. Hoogerland, K. Baldwin, and S. Buckman, Appl. Opt. {\bf
40},  1907 (2001).

\bibitem{Hoogerland3}
R. Dall, M. Hoogerland, K. Baldwin, and S. Buckman, J. Opt. B {\bf 1},
396 (1999)

\bibitem{Rooijakkers}
W. Rooijakkers, W. Hoogervorst, and W. Vassen, Opt. Commun. {\bf 123},
321   (1996)

\bibitem{Herschbach}
N. Herschbach, P. Tol, W. Hogervorst, and W. Vassen, Phys. Rev. A {\bf
61},   050702  (2000)

\bibitem{Tol}
P. Tol {\it et~al.}, Phys. Rev. A {\bf 60},  R761  (1999)

\bibitem{Leduc}
E. Rasel {\it et~al.}, Eur. Phys. J. D {\bf 7},  311  (1999)

\bibitem{Labeyrie}
G. Labeyrie {\it et~al.}, Eur. Phys. J. D {\bf 7},  341  (1999)

\bibitem{Nowak}
S. Nowak {\it et~al.}, Appl. Phys. B {\bf 70},  455  (2000)

\bibitem{HalPeter}
H. Metcalf and P. van~der Straten, {\em Laser cooling and trapping}
(Springer,  New York, 1999).

\bibitem{Hal}
B. Sheehy, S. Shang, P. van~der Straten, and H. Metcalf, Chem. Phys. {\bf
145},    317  (1990).

\bibitem{Phillips}
W. Phillips and H. Metcalf, Phys. Rev. Lett. {\bf 48},  596  (1982)

\bibitem{Lison}
F. Lison, P. Schuh, D. Haubrich, and D. Meschede, Phys. Rev. A {\bf 61},
  013405  (1999).

\bibitem{Hotop}
H. Hotop,  in {\em Atoms and Molecules}, Vol.~29b of {\em Atomic,
Molecular and
  Optical Physics} (Academic Press, New York, 1996), p.\ 191.

\bibitem{Kuppens}
S. Kuppens {\it et~al.}, Phys. Rev. A, submitted for publication

\bibitem{Stas}
R. Stas, Master's thesis, Eindhoven University of Technology, 1999
  (unpublished)

\bibitem{twosnote}
In the formulas we give here several factors of 2 appear compared to
those in
  Ref.~\protect{\cite{Lison}} because we define spatial and velocity spreads in
  terms of their FWHM rather than HWHM values.

\bibitem{rubberbible}
D.~R.~Lide (ed), {\em CRC Handbook of Chemistry and Physics}, 72nd ed.
(CRC Press,
  Boca Raton, 1991).

\bibitem{Laloe}
L. Julien, M. Pinard, and F. Laloe, J. Phys. (Paris) Lett. {\bf 41},  L479
  (1980)

\bibitem{Shimizu1}
F. Shimizu, K. Shimizu, and H. Takuma, Phs. Rev. A {\bf 39},  2758 (1989)

\bibitem{Robert}
A. Robert {\it et~al.}, Science {\bf 292},  461  (2001)

\bibitem{Pereira3}
F.~D. Santos {\it et~al.}, Eur. Phys. J. AP {\bf 14},  69  (2001)

\bibitem{Hoogerland4}
M. Hoogerland, private communication.

\end{thebibliography}

%\onecolumn
\newpage

\begin{table}
\caption{Axial starting position $z$ and length $\Delta z$ of beam line
components, with laser beam characteristics for the different cooling
stages at a total laser output of 700~mW: total optical power used $P$,
average on-resonance saturation parameter $\langle s_0 \rangle$ and laser
detuning $\Delta_L$.}
\begin{tabular}{lccccc}
 Component&$z$&$\Delta z$ &$P$&$\langle s_0 \rangle$ \tablenotemark[1] &$\Delta_L/\Gamma$\\
 \ &(mm)&(mm)&(mW)&\ &\\
 \hline
 nozzle source&0\\
 collimator&43&150&$4\times 20$\tablenotemark[2]&1.6&+8\\
 1$^{\rm{st}}$ wire scanner&210&\\
 1$^{\rm{st}}$ Doppler cooler&230&50&50&3.0&-1.8\\
 beam stop&300&\\
 flow restriction&390&100\\
 2$^{\rm{nd}}$ wire scanner&790&\\
 1$^{\rm{st}}$ Zeeman solenoid&1230&850&100&2.6&-50\\
 2$^{\rm{nd}}$ Doppler cooler&2260&40&40&0.5&-1.8\\
 2$^{\rm{nd}}$ Zeeman solenoid&2370&150&100&2.6&-50\\
 MOC&2670&90&250&1.9&-1.8\\
 3$^{\rm{rd}}$wire scanner&2800&\\
 4$^{\rm{th}}$wire scanner&2950&\\
 slower mirror&3000&\\
 beam flux detector & 3320\\
\end{tabular}
 \tablenotetext[1]{For the Zeeman slower and MOC the light
is circularly polarized while for the other stages it is linearly
polarized}
 \tablenotetext[2]{The collimator
laser beam is split in two pairs of laser beams, one pair for cooling in
the $x$ direction and one pair for cooling in the $y$ direction.}
\end{table}

\begin{table}
\caption{Characteristics of the atomic beam. The total flux $\dot{N}$,
FWHM diameter $d$, average longitudinal velocity $v_{||}$ and FWHM
transverse velocity spread $\Delta v_{\perp}$, brightness ${\cal{R}}$,
brilliance ${\cal{B}}$ and reduced phase-space density
${\tilde{\Lambda}}$ are given as measured at position $z_m$ (second
column) when the setup is operational up to the given cooling stage. The
last three columns give the brightness, brilliance and phase-space
density measured at the position of the flux detector ($z=3320$ mm) when
the setup is operational up to the given cooling stage, relative to the
values for full operation.}
\begin{tabular}{l|cccccccc|ccc}
 stage&$z_m$&$\dot{N}$&$d$&$\bar{v}_{||}$&$\Delta v_{\perp}$&${\cal{R}}$&${\cal{B}}$&$\tilde{\Lambda}$
 &${\cal{R}}/{\cal{R}}_{\rm{All}}$ &${\cal{B}}/{\cal{B}}_{\rm{All}}$ &$\tilde{\Lambda}/\tilde{\Lambda}_{\rm{All}}$\\
 &  &           &  &   &   &$10^{20}/$   &$10^{20}/$   &         \\
 &mm&$10^{10}$/s&mm&m/s&m/s&(s m$^{2}$sr)&(s m$^{2}$sr)&10$^{-13}$\\
 \hline &&&&&&&&&&&\\
 source&0&2600&0.15&480&580&130&1250&580 &4$\times 10^{-7}$&2$\times 10^{-7}$&3$\times 10^{-10}$\\
 collimator&210&110&9&480&11&0.42&4.0&1.9 &1$\times 10^{-2}$&5$\times 10^{-3}$&9$\times 10^{-6}$\\
 1$^{\rm{st}}$ Doppler cooler&790&110&15&480&4.8&0.79&7.6&3.5& 3$\times 10^{-1}$&1$\times 10^{-1}$&2$\times 10^{-4}$\\
 Zeeman slower&2800&55\tablenotemark[1]&30\tablenotemark[1]&98&2.0\tablenotemark[1]&0.01&0.26&70& 1$\times 10^{-2}$&1$\times 10^{-2}$&1$\times 10^{-2}$\\
 MOC&2800&4.8&0.72&98&0.84&21&500&1.3$\times 10^{6}$ &1&1&1\\
\end{tabular}
 \tablenotetext[1]{approximate values}
 \end{table}

\newpage

\newpage
%\twocolumn

\newpage

\newpage

\newpage

\newpage

\newpage

\newpage

\newpage

\end{document}